\documentstyle[12pt]{article}
 
\begin{document}

%%%%%%%%%%%%%%%%%%%%%%%%%%%%%%%%%%%%%%%%%%%%%%%%%%%%%%%%%%%%%%%%%%%%%%%%

\title{ Third Quantization of Brans-Dicke Cosmology }

%%%%%%%%%%%%%%%%%%%%%%%%%%%%%%%%%%%%%%%%%%%%%%%%%%%%%%%%%%%%%%%%%%%%%%%%
\author{Luis O. Pimentel\footnote{lopr@xanum.uam.mx}\,\,\ddag \quad and \quad C\'esar  Mora\footnote{ceml@xanum.uam.mx}\,\,\ddag\S\\ 
\ddag Departamento de F\'{\i}sica ,\\
Universidad Aut\'onoma Metropolitana,\\
Apartado Postal 55-534,CP 09340 M\'exico D.F.,MEXICO.\\
\S Departamento de F\'{\i}sica ,\\
UPIBI-Instituto Polit\'ecnico Nacional,\\
Av. Acueducto s/n Col. Barrio La Laguna Ticom\'an, \\
CP 07340 M\'exico DF, MEXICO. \\}

\date{\today}

\maketitle

%%%%%%%%%%%%%%%%%%%%%%%%%%%%%%%%%%%%%%%%%%%%%%%%%%%%%%%%%%%%%%%%%%%%%%%%%% 

\begin{abstract}                         

%%%%%%%%%%%%%%%%%%%%%%%%%%%%%%%%%%%%%%%%%%%%%%%%%%%%%%%%%%%%%%%%%%%%%%%%%% 
We study the third quantization of a Brans-Dicke toy model, we    calculate the number density of the universes created from nothing and found that it has a Planckian form. Also, we calculated the uncertainty relation for this model by means of functional Schr\"odinger equation and we found that fluctuations of the third-quantized universe field tends to a finite limit in the course of cosmic expansion.

\end{abstract}
Pacs: 98.80.Hw, 04.60.-m, 04.20.Ch.  
 
%%%%%%%%%%%%%%%%%%%%%%%%%%%%%%%%%%%%%%%%%%%%%%%%%%%%%%%%%%%%%%%%%%%%%%%%%% 

\section{Introduction}                         

%%%%%%%%%%%%%%%%%%%%%%%%%%%%%%%%%%%%%%%%%%%%%%%%%%%%%%%%%%%%%%%%%%%%%%%%%% 
  
Quantum cosmology is a consequence of the quantum theory of the universe as a whole and is conventionally based on the Wheeler-DeWitt (WDW) equation\cite{1} $H\psi=0$, where $\psi$ is a wavefunction that describes the universe. This equation is obtained by means of canonical quantization of Hamiltonian $H$. It has problems with the probabilistic interpretation. In the usual formulation of quantum mechanics a conserved positive-definite probability density is required for a consistent interpretation of the physical properties of a given system, and the universe in the quantum cosmology perspective, does not satisfy this requirement, because the WDW equation is a hyperbolic second order differential equation, there is no conserved positive-definite probability density as in the case of the Klein-Gordon equation. An alternative to this problem, is to regard the wavefunction as a quantum field in minisuperspace rather than a state amplitude, and the strategy is to perform a third quantization in analogy with the quantum field theory \cite{2}. Then third-quantized universe theory describes a system of many universes.

When we follows backwards in time the evolution of the universe, the curvatures and densities approach the Planck scale, where one would expect quantum gravitational effects to become important, thus the description based on third-quantized universe theory is of importance at the early moments of the evolution of the universe.

In this paper we  examine the third quantization of a Brans-Dicke solvable model\cite{3} for a closed Friedmann-Robertson-Walker (FRW) universe. The paper is organized as follows. In Sec. II we review the minisuperspace toy model, Sec. III is devoted to study the third quantization of our toy model, in Sec. IV we calculate  the number density of universes created from nothing. We calculate the fluctuations of the third quantized universe in Sec. V, and finally we summarize and conclude in Sec. VI.
%%%%%%%%%%%%%%%%%%%%%%%%%%%%%%%%%%%%%%%%%%%%%%%%%%%%%%%%%%%%%%%%%%%%%%%%%%%

\section {Quantum Cosmology of Brans-Dicke Toy Model}

%%%%%%%%%%%%%%%%%%%%%%%%%%%%%%%%%%%%%%%%%%%%%%%%%%%%%%%%%%%%%%%%%%%%%%%%%%%  
\noindent
Our starting point is the  action for Brans-Dicke theory:
\begin{equation}
S=\frac{1}{l_p^2}\int\!\sqrt{-g}\left[\phi R^{(4)} -
\frac{\omega_0}{\phi} g^{\mu\nu}\phi_{,\mu}\phi_{,\nu}\right]\,d^4\!x,
\label{action0}
\end{equation}
where $\phi(t)$ is the conventional real scalar gravitational field, $l_p$ is the Planck length and $R^{(4)}$ is the curvature constant. We shall consider an homogeneous and isotropic cosmological model described by the  FRW metric
\begin{equation}
ds^2=\sigma^2[N^2(t)dt^2 - a^2(t)d\Omega_3^2],
\label{metric}
\end{equation}
where $N(t)$ is the lapse function, $a(t)$ is the scale factor of the universe, $\sigma^2=l_P^2/24\pi^2$, and $d\Omega^2_3$ is the metric on the unit 3-sphere. By means of new variables and conformal time
\begin{equation}
x=\ln(a^2\phi) \qquad , \qquad y=\ln\phi^{\frac{1}{\rho}}\qquad ,\qquad dt=ad\tau,
\label{var-1}
\end{equation}
where $\rho^2=\frac{3}{2\omega_0+3}$ and choosing $N(t)=1$, action (\ref{action0}) takes the simple form
\begin{equation}
S = \frac{1}{2}\int\left[\frac{x^{\prime2}}{4} - 
\frac{y^{\prime2}}{4} - 1 \right]e^xd\,\tau.
\label{action2}
\end{equation}
Hence we get the Hamiltonian ${H}$ of the system 
\begin{equation}
H = 2e^{-x}\pi_x^2 - 2e^{-x}\pi_y^2
+ \frac{e^x}{2},
\label{hamiltonian0}
\end{equation}
the temporal evolution of dynamical variables is generated by (\ref{hamiltonian0}). The WDW equation is obtained by means of canonical quantization of Hamiltonian constraint $H\Psi(x,y)=0$
\begin{equation}
\left[x^{-\xi}\frac{\partial}{\partial x}\left(x^{\xi}\frac{\partial}{\partial x}\right) - 
\frac{\partial^2}{\partial y^2} - \frac{e^{2x}}{4}\right]\psi(x,y)=0,
\label{wdw1}
\end{equation}
the ambiguity of factor ordering is encoded in the $\xi$ parameter, and $\psi(x,y)$ is the universe wavefunction. The solution to this equation for $\xi=0$, is
\begin{equation}
\psi_p(x,y) = Q_pZ_{\pm\delta}(w)e^{ipy},
\label{sol1}
\end{equation}
where $w=ie^x/2$, $Q_p$ is a normalization factor, $Z_\delta$ is a Bessel function, $\delta=-i\mid p\mid$, and $p$ is an arbitrary separation constant.
 
%%%%%%%%%%%%%%%%%%%%%%%%%%%%%%%%%%%%%%%%%%%%%%%%%%%%%%%%%%%%%%%%%%%%%%%%%%%

\section{Third quantization theory}
 
%%%%%%%%%%%%%%%%%%%%%%%%%%%%%%%%%%%%%%%%%%%%%%%%%%%%%%%%%%%%%%%%%%%%%%%%%%%
\noindent 
It is known that the WDW equation is a result of the quantization of a geo\-me\-try and mater (called second quantization of gravity). The third quantization procedure is obtained by considering the wavefunction   $\psi(x,y)$ as an operator acting on the state vectors of a system of universes and can be decomposed as
\begin{equation}
\psi(x,y)=\int_{-\infty}^{\infty}\left[ C(p)\psi_p(x,y) + 
C^{\dag}(p)\psi_p^*(x,y)\right]dp,
\label{comb}
\end{equation}
where $\psi(x,y)$ and $\psi^*(x,y)$ form complete orthonormal sets of solutions to the WDW equation. This is in analogy with the quantum
field theory, where $C(p)$ and $C^{\dag}(p)$ are creation and 
annihilation operators, we assume that the creation and annihilation operators of universes obey the standard commutation relations
\begin{equation}
  [C(p), C^{\dag}(q)]= \delta(p-q)\quad ,\quad
  [C(p), C(q)]=[C^{\dag}(p), C^{\dag}(q)]=0\quad.
\label{comm1}
\end{equation}
The universe Fock space is spanned by $C^{\dag}(p_1)C^{\dag}(p_2)...\mid 0 >$, and the ground state, {\em i.e.}, the vacuum state $\mid 0 >$ is defined by $C(p)\mid 0 > = 0 \quad \hbox{for}\quad \forall p$.
We will define the initial vacuum state with respect to the orthonormal set of mode solutions which are positive frequency modes in the limit of vanishing spatial volume\cite{4}. Since equation (\ref{wdw1}) is of the Klein-Gordon type, we take variable $x$ as time, and variable $y$ as space. The wavefunction $\psi(x,y)$ is interpreted as a quantum field in minisuperspace. The third-quantized action to yield the WDW equation (\ref{wdw1}) is
\begin{equation}
S_{3Q} = \frac{1}{2}\int\left[\left(\frac{\partial\psi}{\partial x}\right)^2 - \left(\frac{\partial\psi}{\partial y}\right)^2 + \frac{e^{2x}}{4}\psi^2\right]dx\,dy,
\label{action3}
\end{equation}
then the third-quantized Lagrangian is
\begin{equation}
L_{3Q} = \frac{1}{2}\left[\left(\frac{\partial\psi}{\partial x}\right)^2 - \left(\frac{\partial\psi}{\partial y}\right)^2 + \frac{e^{2x}}{4}\psi^2\right].
\label{Lagrangian1}
\end{equation}
The canonical momentum is given by
\begin{equation}
\pi(x,y) = \frac{\partial L_{3Q}}{\partial\frac{\partial\psi}{\partial x}} =\frac{\partial\psi}{\partial x},
\label{momentum}
\end{equation}
and the third-quantized Hamiltonian is
\begin{equation}
H_{3Q} = \frac{1}{2}\left[\left(\frac{\partial\psi}{\partial x}\right)^2 + \left(\frac{\partial\psi}{\partial y}\right)^2 + \frac{e^{2x}}{4}\psi^2\right].
\label{Hamiltonian1}
\end{equation}
Now, in order to quantize this toy model, we impose the equal time commutation relations
\begin{equation}
\left[i\frac{\partial\psi(x,y)}{\partial x}, \psi(x,y^\prime)\right]=
\delta(y-y^\prime),\,\  \left[i\frac{\partial\psi(x,y)}{\partial x}, i\frac{\partial\psi(x,y^\prime)}{\partial x}\right] = \left[\psi(x,y), \psi(x,y^\prime)\right] =0.
\end{equation}
We can find a complete set of orthonormal positive frequency solutions to WDW equation (\ref{wdw1}) in terms of the following inner product
\begin{equation}
<\psi_p|\psi_{q}>=\delta(p-q),\quad <\psi_p^*|\psi_{q}>=0,\quad <\psi_p^*|\psi_q^*>=-\delta(p-q),
\end{equation}
where the Klein-Gordon scalar product is defined as
\begin{equation}
<\psi_p,\psi_q>=
i\int \psi_p\stackrel{\leftrightarrow}{\partial}_x
\psi_q^*dy = \delta(p-q).
\label{inner1}
\end{equation}
By means of these normal modes $\psi_p$, we expand the field $\psi(x,y)$ in terms of equation (\ref{comb}).

%%%%%%%%%%%%%%%%%%%%%%%%%%%%%%%%%%%%%%%%%%%%%%%%%%%%%%%%%%%%%%%%%%%%%%%%

\section{ Universe Creation from Nothing }

%%%%%%%%%%%%%%%%%%%%%%%%%%%%%%%%%%%%%%%%%%%%%%%%%%%%%%%%%%%%%%%%%%%%%%%%
\noindent
In order to third quantize the universe, we can construct the normal mode functions $\psi_p^{in}(x,y)$ and $\psi_p^{out}(x,y)$ from (\ref{sol1}), in this way, we chose the in-mode function as a normalized function that satisfies (\ref{inner1}),
\begin{equation}
\psi_p^{in}(x,y) = Q_p^{in}J_\delta(w)e^{ipy},
\label{modein1a}
\end{equation}
where the normalization factor is
\begin{equation}
 Q_p^{in}= \frac{e^{\frac{\pi}{2}\mid p\mid}}{2\sinh^{\frac{1}{2}}\left(\pi \mid p\mid \right)},
\label{const-modein1}
\end{equation}
in this solution the separation constant $p$ can be regarded as a canonical momentum of $y$. When $x \to -\infty$, the asymptotic behavior of (\ref{modein1a}) is
\begin{equation}
\psi_p^{in}(x,y)\approx e^{-i\left(\mid p \mid x - py\right)},
\label{modein1b}
\end{equation}
and (\ref{modein1a}) is a positive frequency mode function. In the classically allowed regions the positive frequency modes correspond to the expanding universe\cite{5}. The out-mode function is defined by
\begin{equation}
\psi_p^{out}(x,y) = Q_p^{out}H_{-\delta}^{(2)}(w)e^{ipy},
\label{modeout1a}
\end{equation}
where
\begin{equation}
Q_p^{out} = \frac{1}{2\sqrt2}e^{-\frac{\pi}{2}\mid p\mid},
\label{const-modeout1a} 
\end{equation}
when we take the limit $x \to \infty$, (\ref{modeout1a}) tends to
\begin{equation}
\psi_p^{out}(x,y) \approx e^{-\frac{x}{2}-i\left(\frac{e^{x}}{2}-py\right)},
\label{modeout1b}
\end{equation}
the above equation is a positive frequency out-mode function for large scales.  The expansion of $\Psi(x,y)$ in terms of creation and annihilation operators for the in-mode function is 
\begin{equation}
\psi(x,y)=\int\left[C_{in}(p)\psi_p^{in}(x,y) +
C_{in}^\dag(p)\psi_p^{in*}(x,y)\right]dp.
\label{expan-in}
\end{equation}
We require that at very early times the vacuum be annihilated by $C_{in}(p)$, then the in-vacuum $|0,in>$, which is regarded as "nothing" is defined by
\begin{equation}
C_{in}(p)|0,in>=0  \quad \hbox{for p} \in {\bf R},
\end{equation}
also the expansion of $\psi(x,y)$ for the out-mode function is
\begin{equation}
\psi(x,y)=\int\left[C_{out}(p)\psi_p^{out}(x,y) +
C_{out}^\dag(p)\psi_p^{out*}(x,y)\right]dp,
\label{expan-out}
\end{equation}
and the out vacuum $|0,out>$ is defined as
\begin{equation}
C_{out}(p)|0,out>=0  \quad \hbox{for p} \in {\bf R}.
\end{equation}
Now, we will proceed to obtain the average number of universes produced from nothing by calculating the Bogoliubov transformation coefficients between $in$ and $out$ fields. Since both sets (\ref{modein1a}) and (\ref{modeout1a}) are complete, they are related to each other by a Bogoliubov transformation,  
\begin{equation}
\psi_p^{out}(x,y)=\int\left[C_1(p,q)\psi_q^{in}(x,y) +
C_2(p,q)\psi_q^{in*}(x,y)\right]dq.
\end{equation} 
We obtained that the Bogoliubov coefficients $C_\eta(p,q)=\delta(p\mp q)C_\eta$ are
\begin{equation}
C_1(p,q) = \delta(p-q)\frac{1}{\sqrt{1-e^{-2\pi\mid p\mid}}},
\end{equation}
and
\begin{equation}
C_2(p,q) = \delta(p+q)\frac{1}{\sqrt{e^{2\pi\mid p\mid}-1}}.
\end{equation}
The coefficients $C_1(p)$ and $C_2(p)$ are not equal to zero and satisfy the probability conservation condition $\mid C_1(p)\mid^2 - \mid C_2(p)\mid^2=1$. Thus, two Fock spaces constructed with the help of the modes (\ref{modein1a}) and (\ref{modeout1a}) are not equivalent and we have two different third quantized vacuum states (voids): The in-vacuum $\mid 0, in\rangle$ and out-vacuum $\mid 0, out\rangle$, defined by
\begin{equation}
  C_{in}(p)\mid 0, in \rangle  = 0
    \quad \hbox{and} \quad
    C_{out}(p)\mid 0, out \rangle = 0,
 \end{equation}
where $p \in {\bf R}$. The average number of universes produced from nothing $\em{i.e.}$, the in-vacuum, in the 
p-th mode $N(p)$ is
\begin{eqnarray}
  N(p)&=&\left<0,in\mid C_{out}^{\dag}(p)C_{out}(p)
  \mid 0, in \right>=\mid C_2(p)\mid^2\nonumber\\
      &=& \frac{1}{e^{2\pi\mid p\mid}-1}.
\end{eqnarray}
We can see that this is a Planckian distribution with respect to $\mid p\mid$ for a massless gas in 1+1 dimensions, $\em{i.e.}$, the initial state $\mid 0, in>$ is populated by a thermal distribution of universes.  

%%%%%%%%%%%%%%%%%%%%%%%%%%%%%%%%%%%%%%%%%%%%%%%%%%%%%%%%%%%%%%%%%%%%%%%%%

\section{Uncertainty Relation}

%%%%%%%%%%%%%%%%%%%%%%%%%%%%%%%%%%%%%%%%%%%%%%%%%%%%%%%%%%%%%%%%%%%%%%%%%
\noindent 
In this section we will use a Fourier decomposition in order to obtain decoupled degrees of freedom, the universe field $\psi(x,y)$ will be expand in sine and cosine functions. We assume that our quantum system is confined to a one-dimensional box with periodic boundary conditions, with the coordinate length fixed at a arbitrary value $M$, then
\begin{equation}
\psi(x,y)=\frac{\sqrt{2\pi}}{M}\left\{\psi(x,0)+\sum_{q=2\pi n/M} \frac{1}{\sqrt{2}}
\left[\psi_+(x,q)\cos{qy}+\psi_-(x,q)\sin{qy}\right]\right\},
\label{Fourier}
\end{equation}
where $\psi_+(x,-q)=\psi_+(x,q)$ and $\psi_-(x,-q)=\psi_-(x,q)$. By substituting (\ref{Fourier}) into the Lagrangian (\ref{Lagrangian1}) with the variables 
\begin{equation}
z=a^2\phi, \qquad \hbox{and}\qquad y=\rho^{-1}\ln{\phi}
\label{var-2}
\end{equation}
we obtain
\begin{equation}
L=\frac{1}{2}\sum_\alpha\left[\left(\frac{\partial\psi_\alpha(z)}{\partial z}\right)^2-q^2\psi_\alpha^2(z)+\frac{z^2}{4}\psi_\alpha^2(z)\right],
\label{Lagrangian2}
\end{equation}
where we have denoted the mode variables $\psi(z,0)$ and $\psi_\pm(z,q)$ by $\psi_\alpha(z)$ and (\ref{Lagrangian2}) is rescaled to $\psi_\alpha \to \sqrt{M/2\pi}\psi_\alpha$. The above sum includes zero mode $\psi(z,0)$ for each pair $(q,-q)$, in this way the mode variable $\psi_\alpha$ is completely decoupled from each other. Now the Hamiltonian is given by
\begin{equation}
H_\alpha=\sum_\alpha\frac{1}{2}\left[\pi_\alpha^2+\left(q^2-\frac{z^2}{4}
\right)\psi_\alpha^2(z)\right].
\label{Hamiltonian2}
\end{equation}
To quantize this system (\ref{Hamiltonian2}), we impose the equal time commutation relations
\begin{equation}
\left[\hat\psi_\alpha(z),\hat\pi_\alpha^\prime\right]=i\delta_{\alpha,
\alpha^\prime}.
\end{equation}
The wave functional of any energy eigenstate is factorized as
\begin{equation}
\Psi = \prod_\alpha\Psi_\alpha\left[z,\psi_\alpha(z)\right].
\end{equation}
In order to get the Heisenberg uncertainty relation we use the Schr\"odinger picture, then we make the substitution $\hat\psi(z) \to \psi_\alpha(z)$ and $\hat\pi_\alpha \to -i\frac{d}{d\psi_\alpha(z)}$. 
Then, the Schr\"odinger equation for each mode variable is
\begin{equation}
i\frac{\partial \Psi_\alpha}{\partial z} = \frac{1}{2}\left[-\frac{\partial^2}{\partial\psi_\alpha^2} + \left(q^2-\frac{z^2}{4}\right)\psi_\alpha^2\right]\Psi_\alpha.
\label{Sch}
\end{equation}
We will solve the above wave functional equation by using the Gaussian ansatz
\begin{equation}
\Psi_\alpha[z,\psi_\alpha] = C\exp\{-\frac{1}{2}A_\alpha(z,q)[\psi_\alpha(z,q)-\eta_\alpha(z,q)]^2 + iP_\alpha(z,q)[\psi_\alpha(z,q)-\eta_\alpha(z,q)]\},
\label{ansatz}
\end{equation}

\begin{equation}
A_\alpha(z,q) = D_\alpha(z,q) + iF_\alpha(z,q),
\label{deff}
\end{equation}
where the real functions $D_\alpha(z,q)$, $F_\alpha(z,q)$ $P_\alpha(z,q)$, and $\eta_\alpha(z,q)$ have to be determined from equation (\ref{Sch}). $C$ is a normalization of the wave function. The inner product of two functionals $\Psi_{1\alpha}(z,\psi)$ and $\Psi_{2\alpha}(z,\psi)$ is defined by
\begin{equation}
<\Psi_1\mid \Psi_2>_z = \int d\psi_\alpha\Psi_{1\alpha}[z,\psi_\alpha]
\Psi^*_{2\alpha}[z,\psi_\alpha].
\label{inner2}
\end{equation}
We shall calculate Heisenberg's uncertainty relation, the dispersion of $\psi_\alpha$ is given by $\left(\Delta\psi_\alpha\right)^2 \equiv <\psi_\alpha^2>_z - <\psi_\alpha>^2_z,$ from equations (\ref{ansatz}) and (\ref{inner2}) we have
\begin{equation}
<\psi_\alpha^2>_z = \frac{1}{2D_\alpha(z,q)} + \eta^2_\alpha(z,q), \quad
<\psi_\alpha>_z = \eta^2_\alpha(z,q),
\end{equation}
then 
\begin{equation}
\left(\Delta\psi_\alpha\right)^2 = \frac{1}{2D_\alpha(z,q)},
\end{equation}
and the dispersion of $\pi_\alpha$ is  $\left(\Delta\pi_\alpha\right)^2 \equiv  <\pi_\alpha^2>_z - <\pi_\alpha>^2_z$, where
\begin{equation}
<\pi_\alpha^2>_z = \frac{D_\alpha(z,q)}{2} + 
\frac{F^2_\alpha(z,q)}{2D_\alpha(z,q)} + P^2_\alpha(z,q), \quad
<\pi_\alpha>_z = P_\alpha(z,q),
\end{equation}
then simplifying
\begin{equation}
\left(\Delta\pi_\alpha\right)^2 = \frac{D_\alpha(z,q)}{2} + 
\frac{F^2_\alpha(z,q)}{2D_\alpha(z,q)},
\end{equation}
thus we obtain the uncertainty relation
\begin{equation}
\left(\Delta\psi_\alpha\right)^2\left(\Delta\pi_\alpha\right)^2 = \frac{1}{4}
\left\{1+\left[\frac{F_\alpha(z,q)}{D_\alpha(z,q)}\right]^2\right\}.
\label{rel-uncertainty}
\end{equation}
In order to obtain $ F_\alpha(z,q)$ and $D_\alpha(z,q)$ we substitute the ansatz (\ref{ansatz}) into equation (\ref{Sch}), thus we obtain the equation of motion for $A_\alpha(z,q)$:
\begin{equation}
-i\frac{d}{dz}A_\alpha(z,q) = -A^2_\alpha(z,q) + q^2 - \frac{z^2}{4}.
\end{equation}
Now we substitute
\begin{equation}
A_\alpha(z,q) = -i\frac{d}{dz}\ln{u_\alpha(z,q)} ,
\label{mot}
\end{equation}
where $u_\alpha(z,q)$ is the solution of the WDW equation
\begin{equation}
\left[\frac{\partial^2}{\partial z^2} + \frac{q^2}{z^2} - \frac{1}{4}\right]u_\alpha(z,q) = 0.
\label{wdw-2}
\end{equation}
First, we analyze the behavior of $u_\alpha$ for large scales $(a\to\infty)$, then the second term  $q^2/z^2$ in the wave equation (\ref{wdw-2}) can be neglected. Thus, the solution is
\begin{equation}
u_\alpha(z,q)=\cos{\frac{z}{2}} + \beta \sin{\frac{z}{2}},
\label{wdw-2-solut}
\end{equation}
where $\beta$ is a complex constant. Substituting (\ref{wdw-2-solut}) into  (\ref{mot}), we obtain that
\begin{equation}
A_\alpha(z,q)=\frac{i}{2}\frac{\sin{\frac{z}{2}}-\beta\cos{\frac{z}{2}}}
{\left(\cos{\frac{z}{2}}+\beta\sin{\frac{z}{2}}\right)},
\label{eq-A}
\end{equation}
then
\begin{equation}
D_\alpha(z,q)=\hbox{Re}A_\alpha(z,q)= 
\frac{\hbox{Im}\beta}{2\mid\cos{\frac{z}{2}}+\beta\sin{\frac{z}{2}}\mid^2},
\end{equation}
and
\begin{equation}
F_\alpha(z,q)=-\hbox{Im}A_\alpha(z,q)=
\frac{\hbox{Re}\beta\cos{z}-\frac{1}{2}(\mid\beta\mid^2-1)\sin{z}}
{2\mid\cos{\frac{z}{2}}+\beta\sin{\frac{z}{2}}\mid^2},
\end{equation}
therefore the uncertainty relation becomes
\begin{equation}
\left(\Delta\psi_\alpha\right)^2\left(\Delta\pi_\alpha\right)^2 = \frac{1}{4}
\left\{1+(\hbox{Im}\beta)^{-2}\left[\hbox{Re}\beta\cos{z}-\frac{1}{2}(\mid\beta\mid^2-1)\sin{z}\right]^2\right\}.
\label{end-uncertainty-1}
\end{equation}
If we chose $\beta=-i$, then for large scales $(a\to\infty)$ the above uncertainty relation becomes
\begin{equation}
\left(\Delta\psi_\alpha\right)^2\left(\Delta\pi_\alpha\right)^2 \simeq \frac{1}{4},
\label{uncertainty-1}
\end{equation}
for $\beta\neq-i$ and $\mid\beta \mid << 1$, we get that 
\begin{equation}
\left(\Delta\psi_\alpha\right)^2\left(\Delta\pi_\alpha\right)^2 \simeq \frac{1}{4}\left[1+O^{-1}(\mid\beta\mid^2)\right],
\label{uncertainty-2a}
\end{equation}
and when $\mid \beta \mid >> 1$
\begin{equation}
\left(\Delta\psi_\alpha\right)^2\left(\Delta\pi_\alpha\right)^2 \simeq \frac{1}{4}\left[1+O(\mid\beta\mid^2)\right].
\label{uncertainty-2b}
\end{equation}
This means that the quantum fluctuations of the third-quantized closed universe field are bounded at a finite value according equation (\ref{end-uncertainty-1}), in the course of the universe expansion $(a\to \infty)$. Now we study the behavior of $u_\alpha$ for small scales $(a\to 0)$. The solution to equation (\ref{wdw-2}) is
\begin{equation}
u_\alpha(z,q)=\sqrt{z}\left[I_\nu\left(\frac{z}{2}\right)+\gamma K_\nu\left(\frac{z}{2}\right)\right],
\label{wdw-2-sol}
\end{equation}
where $I_\nu$ and $K_\nu$ are modified Bessel functions, $\gamma$ is a complex constant, and $\nu=\sqrt{1/4-q^2}$. Substituting the general solution (\ref{wdw-2-sol}) into (\ref{mot}), we obtain
\begin{equation}
D_\alpha=\hbox{Re}A_\alpha=\frac{\hbox{Im}\gamma(K_\nu^\prime I_\nu-I_\nu^\prime K_\nu)}{2\mid I\nu+\gamma K_\nu\mid^2},
\end{equation}
\begin{equation}
F_\alpha=\hbox{Im}A_\alpha 
= -\frac{z^{-1}\mid I_\nu+\gamma K_\nu\mid^2 + I^\prime_\nu I_\nu+
\mid\gamma\mid^2K^\prime_\nu K_\nu + \hbox{Re}\gamma(K^\prime_\nu I_\nu
+I^\prime_\nu K_\nu)}{2\mid I\nu+\gamma K_\nu\mid^2},
\end{equation}
the prime denotes differentiation with respect to $z$. Then, the uncertainty relation is
\begin{equation}
(\Delta\psi_\alpha)^2(\Delta\pi_\alpha)^2=
\frac{1}{4}\left\{1+G_\alpha^2\right\},
\label{uncert-2f}
\end{equation}
where
\begin{equation}
G_\alpha=(\hbox{Im}\gamma)^{-1}z\left[z^{-1}(I_\nu^2+\mid\gamma\mid^2 K_\nu^2) + I^\prime_\nu I_\nu+
\mid\gamma\mid^2K^\prime_\nu K_\nu + \hbox{Re}\gamma(K^\prime_\nu I_\nu
+I^\prime_\nu K_\nu) \right],
\end{equation}
since we have assumed that our system is confined to a one-dimensional box with periodic boundary conditions, with the coordinate length fixed at an arbitrary value $M$, if we take the limit $M \to \infty$, then we can choose  $\nu^2>0$, and for small scales $(a\to 0)$ the asymptotic behavior\cite{6} of equation  (\ref{uncert-2f}) is
\begin{equation}
(\Delta\psi_\alpha)^2(\Delta\pi_\alpha)^2=
\frac{1}{4}\left\{1+Nx^{-4\mid\nu\mid}\right\},
\end{equation}
where $N$ is some positive constant. This means that the fluctuation of the third-quantized universe field becomes large for small scales $(a\to 0)$. For the case in which $\nu^2<0$, so that $\nu$ is pure imaginary, we obtained an oscillatory behavior of $G_\alpha$ and does not have a definite magnitude\cite{7} when $a \to 0$.

%%%%%%%%%%%%%%%%%%%%%%%%%%%%%%%%%%%%%%%%%%%%%%%%%%%%%%%%%%%%%%%%%%%%%%%%%%

\section{Summary and discussion}

%%%%%%%%%%%%%%%%%%%%%%%%%%%%%%%%%%%%%%%%%%%%%%%%%%%%%%%%%%%%%%%%%%%%%%%%%%
We have studied on the third quantization of a  Brans-Dicke toy model, in which time is related by the scale factor of universe and the space coordinate is related with the scalar field. We have see from this theory that the universe can be created and annihilated according to the law of the field theory. We calculated the number density  of the universes creating from nothing and found that the initial state $\mid 0, in>$ is populated by a  Planckian distribution of baby universes. A related problem of this theory is the detection of universes, this problem has been addressed by analyzing the interaction vertex at which the universe branches off  and introducing an hypothetical internal comoving model detector, which is an analogue of the DeWitt-Unruh detector\cite{7}, which can count the baby universe which plunges into our mother universe\cite{8}. Also, we studied the third-quantization of our model by using the functional Schr\"odinger equation in order to investigate the Heisenberg uncertainty relation and we found that quantum fluctuations of the third-quantized universe field becomes small for large scales  in the course of cosmic expansion. Pohle\cite{9} found a different behavior for the uncertainties, but the reason for that is that he was considering the classically forbidden region pointed by Horiguchi\cite{7}.

%%%%%%%%%%%%%%%%%%%%%%%%%%%%%%%%%%%%%%%%%%%%%%%%%%%%%%%%%%%%%%%%%%%%%%%%%%

\section{Acknowledgments}

%%%%%%%%%%%%%%%%%%%%%%%%%%%%%%%%%%%%%%%%%%%%%%%%%%%%%%%%%%%%%%%%%%%%%%%%%%
\noindent
This work was partially supported by P/FOMES 98-35-15 grant. C M was supported by a EDD-COFAA-IPN grant.

%%%%%%%%%%%%%%%%%%%%%%%%%%%%%%%%%%%%%%%%%%%%%%%%%%%%%%%%%%%%%%%%%%%%%%%%%%

\end{document}